\documentclass[pre,aps,10pt,twocolumn,twoside,floats,preprintnumbers,floatfix,superscriptaddress,a4paper,showpacs,showkeys,longbibliography]%
{revtex4-1}

\usepackage{amsmath}
\usepackage{amsfonts}
\usepackage{amssymb}

\usepackage{hyperref}

\usepackage{times}
\usepackage{time}
\usepackage{graphicx}

\usepackage{color}

\usepackage{lipsum}

 \newcommand{\beq}{\begin{equation}}
 \newcommand{\eeq}{\end{equation}}

   \newcommand{\pder}[2]{\frac{\partial {#1}}{\partial {#2}}}

   \def\rhoRF{\rho_\textrm{f}}
   \def\epsRF{\epsilon_\textrm{f}}
   \def\sRF{s_\textrm{f}}
   \def\pRF{p_\textrm{f}}
   \def\TRF{T_\textrm{f}}

   \def\mi{\mathrm{i}} 

\def\ie{i.e.~}

\def\fig#1{\ref{Fig:#1}}
\def\Fig#1{Fig.~\fig{#1}}
\def\Figs#1{Figs.~\fig{#1}}

\def\eq#1{(\ref{Eq: #1})}
\def\Eq#1{Eq.~\eq{#1}}
\def\Eqs#1{Eqs.~\eq{#1}}

\def\equi{\textrm{eq}}
\def\det{\textrm{det}}
\def\diag{\textrm{diag}}
\def\Re{\textrm{Re}}
\def\Im{\textrm{Im}}

\begin{document}
\title{Stability analysis for a thermodynamically consistent model of relativistic fluid dynamics}

\author{Laura Stricker}
\affiliation{\text{ETH Z\"{u}rich, Department of Materials, Polymer Physics, CH-8093 Z\"{u}rich, Switzerland}}

\author{Hans Christian \"{O}ttinger}
\affiliation{\text{ETH Z\"{u}rich, Department of Materials, Polymer Physics, CH-8093 Z\"{u}rich, Switzerland}}

\date{\today}


\begin{abstract}
The search for thermodynamic admissibility moreover reveals a fundamental difference between liquids and gases in relativistic fluid dynamics, as the reversible convection mechanism is much simpler for liquids than for gases. In relativistic fluid mechanics, positive entropy production is known to be insufficient for guaranteeing stability. Much more restrictive criteria for thermodynamic admissibility have become available in nonequilibrium thermodynamics. We here perform a linear stability analysis for a model of relativistic hydrodynamics that is based on the GENERIC (general equation for the nonequilibrium reversible-irreversible coupling) framework of nonequilibrium thermodynamics. Assuming a simple form of the entropy function, we find stability for the entire range of physically meaningful model parameters. For relativistic fluid dynamics, full thermodynamic admissibility indeed leads to stability.
\end{abstract}

\maketitle

\section{Introduction}
\label{sec:Introduction}
%
%
%
%
%

Relativistic fluid dynamics is important in astrophysics and cosmology as, for example, it allows us to describe the collapse of stars into neutron stars, flows around black holes, jets with relativistic speeds originating from the core of active galactic nuclei, the formation of galaxies or the expansion of the entire universe. The first use of relativistic fluid dynamics in the field of high-energy physics appeared in the 1950s, in the works of Landau \cite{lan53} and Khalatnikov \cite{kha54}. In recent years, the topic came to a renewed attention due to its potential to describe the experiments on quark-gluon plasmas, produced in Relativistic Heavy Ion Colliders (RHIC) \cite{ada05,adc05,ars05,bac05}.

The first attempts to generalize the laws of fluid mechanics and thermodynamics to the relativistic context were based on an extended Fourier law for heat conduction \cite{eck40,lan58_BOOK}. These formulations suffered from two fundamental flaws: they are unstable \cite{HiscockLindblom83,HiscockLindblom85} and the parabolic nature of the classical differential equation for heat flow leads to instantaneous propagation of heat, hence violating causality \cite{kam70}. To overcome this issue, some authors added an \emph{ad hoc} relaxation term to Fourier's law, turning the parabolic equation into a hyperbolic equation with a finite propagation speed \cite{cat58,kra66}.

A big step forward in formulating relativistic fluid dynamics resulted from the insight that the entropy should depend on additional structural variables (related to the momentum and energy fluxes) and that the heat flux should not be proportional to the entropy flux. In 1967, M\"{u}ller demonstrated in the context of heat flow that such a more complete description of nonequilibrium states solves the problem of infinite propagation speeds, and he moreover showed that such a description emerges naturally from kinetic gas theory \cite{mul67}. It was then natural to extend Grad's moment method for the Boltzmann equation \cite{gra49} to the relativistic context \cite{ste71_BOOK,and70,mar69,kra70}. The derivation of moment equations from the relativistic Boltzmann equation culminated in what is now known as the Israel-Stewart model \cite{isr76,isr79}. All work along these lines suffers from the infamous closure problem for the moment hierarchy.

The problem of relativistic fluid dynamics has also been approached with the tools of nonequilibrium thermodynamics. The basic idea of EIT (extended irreversible thermodynamics) is to consider the dissipative fluxes (e.g., the heat flux vector and the viscous pressure tensor) as independent variables and to formulate convection-relaxation equations for such fluxes \cite{jou88}. In the limit of slow phenomena, these equations reduce to the classical constitutive laws, but they are also suitable to describe fast phenomena, since they lead to hyperbolic equations with finite speeds of propagation for thermal and viscous perturbations. Such models are compatible with the kinetic theory of Grad's 14-moment method \cite{bet09}; they are applicable to non-stationary processes, compatible with causality and stable under specific constraints \cite{pu10}. The GENERIC (general equation for the nonequilibrium reversible-irreversible coupling) framework of nonequilibrium thermodynamics is more versatile in the choice of variables than EIT, but more restrictive in the structure of the equations, in particular, by imposing a Hamiltonian structure on reversible dynamics as the basis of a clear and general reversible-irreversible separation. A GENERIC model of relativistic fluid dynamics has been developed in \cite{hco98,hco98b}, and gravity has been included in \cite{hco112,ilg99}. Further possible approaches include Carter's theory \cite{car91}, Van's model \cite{van08} and the conformal field theory formulation \cite{bai08}.

In the present paper we study the linear stability of the GENERIC model developed in \cite{hco98,hco98b}. We first present this model and offer a sketch of the heuristic motivation of these equations for relativistic fluid dynamics offered in the recent textbook \cite{hco18_BOOK}.


\section{Model}
\label{sec:Model}
We restrict ourselves to special relativity. For the metric, we adopt the Minkowski tensor with the convention $(\eta^{\mu\nu}) = \diag(-1,1,1,1)$. For the indexes of four-vectors and tensors, we use Greek letters to denote components from 0 to 3 (\ie time and space components) and Latin letters to denote components from 1 to 3 (\ie spatial components only).
We define a dimensionless velocity four-vector by means of its components
\beq
u^0 = \gamma, \, \, \, u^i = \gamma \frac{v^i}{c} \, \, \, \,    (i = 1,2,3),
\label{Eq: velocity 4-vector}\eeq
where the $v^i$ are the components of the local velocity of the fluid \textbf{v}, $c$ is the speed of light and $\gamma$ is the Lorentz factor, chosen in such a way that
\beq
u^\mu u_\mu = -1,
\label{Eq: Velocity constraint}
\eeq
that is
\beq
\gamma = \frac{1}{\sqrt{1 - \textbf{v}^2/c^2}}.
\label{Eq: Lorentz factor}
\eeq

The relativistic continuity equation reads
\beq
\partial_\mu(\rho_\textrm{f} u^\mu) = 0,
\label{Eq: Continuity eq}
\eeq
with $\rho$ the mass density and the subscript $\textrm{f}$ denoting quantities evaluated in the co-moving local reference frame of the fluid. To define such a frame, different choices are possible: in the Eckart framework, the particle flow vanishes in the local rest frame of the fluid \cite{eck40}, while in the Landau framework, the momentum density vanishes in the local rest frame \cite{lan58_BOOK}. We decide to operate within the Eckart framework, i.e.\ we consider as the local reference frame the frame in which the spatial components of the velocity vanish $u_\textrm{f}^0 = 1, u_\textrm{f}^i = 0$. The tensor $\hat{\eta}^{\mu\nu} = \eta^{\mu\nu} + u^\mu u^\nu$ denotes the spatial projection operator. In the local rest frame such an operator selects the spatial components of vectors and tensors it is applied to.

The energy and momentum balances in the relativistic setting are combined and can be expressed as
\beq
\partial_\mu T^{\mu\nu} = 0,
\label{Eq: Energy-momentum conservation eq}
\eeq
where $T_{\mu\nu}$ is the energy-momentum tensor. Such a tensor has to incorporate energy and momentum densities as well as energy and momentum fluxes.
To avoid the presence of spatial derivatives in the formulation of the fluxes, in the spirit of the description of nonrelativistic complex fluids \cite{hco18_BOOK}, we introduce the relaxing structural variables $\alpha_{\mu\nu}$ and $\omega_\mu$, namely the momentum and energy fluxes respectively. At this point the exact physical meaning of $\alpha_{\mu\nu}$ and $\omega_\mu$ is left open, but it will emerge in the following.
We require
\beq \alpha_{\mu\lambda} u^\lambda = u_\mu.  \label{Eq: Alpha constraint}\eeq
Thus, in the co-moving reference frame, the temporal components are $(\alpha_\textrm{f})_{00} = -1$, $ (\alpha_\textrm{f})_{i0} = (\alpha_\textrm{f})_{0i} = 0$. The four-tensor $\alpha_{\mu\nu}$ is dimensionless and symmetric, therefore it has only six independent components.

By making use of the upper convected Maxwell model for a structural variable \cite{hco18_BOOK}, we express the time relaxation of $\alpha_{\mu\nu}$ as
\beq
u^\lambda (\partial_{\lambda} \alpha_{\mu \nu} - \partial_{\mu} \alpha_{\lambda \nu} -\partial_{\nu} \alpha_{\mu \lambda}) = -\frac{1}{c \lambda_0} \bar{\alpha}_{\mu \nu} - \frac{1}{c \lambda_2} \mathring{\alpha}_{\mu \nu}.
\label{Eq: Relaxation alpha}
\eeq
The left hand side represents the upper convected derivative of $\alpha_{\mu\nu}$ and structurally reminds of a Lie algebra. In the case of a space-like vector, this would be equivalent to the Lie derivative of $\alpha_{\mu\nu}$ in a rigorous sense \cite{mat08}. The upper convected derivative appearing at the left hand side of \Eq{Relaxation alpha} describes the variation in time of the tensor $\alpha_{\mu\nu}$, when following the trajectory of the fluid particle it refers to (translation) as well as the rotation and deformation of $\alpha_{\mu\nu}$ along such a trajectory. In particular, $u^\lambda \partial_{\lambda} \alpha_{\mu \nu}$ is the so-called substantial derivative of $\alpha_{\mu \nu}$ accounting for the time derivative and the convection term, while the remaining two terms on the left-hand side account for rotations and deformations. The parameters $\lambda_0$ and $\lambda_2$ are two constants, representing the characteristic relaxation times of the trace-part $\bar{\alpha}_{\mu \nu}$ and the trace-free part $\mathring{\alpha}_{\mu \nu}$ of $\alpha_{\mu \nu}$,

\beq
\bar{\alpha}_{\mu \nu} = \frac{1}{3} (\alpha_{\lambda}^{\ \lambda} - 1) \hat{\eta}_{\mu\nu}, \, \, \, \, \mathring{\alpha}_{\mu \nu} = \alpha_{\mu \nu} + u_\mu u_\nu - \bar{\alpha}_{\mu \nu}.
\label{Eq: Trace-traceless part alphas}
\eeq
The underlying assumption of \Eq{Relaxation alpha} is that the trace and traceless part of $\alpha_{\mu \nu}$ relax independently from each other.

Similarly, the time relaxation equation for the four-vector $\omega_\mu$ is formulated as
\beq
u^\nu(\partial_\nu \omega_\mu - \partial_\mu \omega_\nu) = -\frac{1}{c\lambda_1} \hat{\eta}_{\mu\nu} \omega^\nu,
\label{Eq: Relaxation omega}
\eeq
where the left-hand side is again an upper convected derivative and $\lambda_1$ is a characteristic relaxation time. From now on, we rename $\hat{\eta}_{\mu\nu} \omega^\nu = \hat{\omega}_\mu$. The main reason why we introduced $\hat{\eta}_{\mu\nu}$ is related to the non-relaxing nature of $\omega_0$ in the local-rest frame, since $(\omega_0)_\textrm{f}$ corresponds to the local equilibrium temperature, as we will show below. The introduction of $\hat{\eta}_{\mu\nu}$ guarantees that both sides of \Eq{Relaxation omega} vanish, upon contraction with $u^\mu$, since
\beq \hat{\omega}_{\mu} u^\mu = 0.  \label{Eq: Omega hat constraint}\eeq

The appropriate constitutive equation for the energy-momentum tensor $T^{\mu\nu}$, is derived from the entropy balance equation, with the assumption that the entropy density in the local rest frame $\sRF$ depends on the mass density $\rhoRF$ and internal energy density $\epsRF$, as well as on the structural variables, \ie it can be expressed as $\sRF (\rhoRF,\epsRF,\alpha_{\mu\nu},\hat{\omega}_\mu)$. The spatial projection $\hat{\omega}_\mu$ is used here instead of $\omega_\mu$ because the temporal part of the latter includes thermodynamic information and would be redundant with $\rhoRF$ and $\epsRF$. From the entropy balance equation, a constraint on the $\omega_\mu$ arises \cite{hco18_BOOK},
\beq
u^\nu \omega_\nu = \TRF,
\label{Eq: Omega constraint}
\eeq
where $\TRF$ is the nonequilibrium temperature in the local rest frame, defined through
\beq
\pder{\sRF}{\epsRF} = \frac{1}{\TRF}.
\label{Eq: Temperature}
\eeq
\Eq{Temperature} is formally similar to the classical definition of thermodynamic temperature in the nonrelativistic setting, but entails some additional ambiguity. In the equilibrium setting, the derivative of $\sRF$ with respect to $\epsRF$ is performed keeping $\rhoRF$ constant. Here, we additionally have to keep the structural variables constant. Therefore the definition of the temperature depends on the particular choice of the structural variables.

The final constitutive equation for the energy-momentum tensor obtained from the entropy balance reads
\begin{eqnarray}
    T^{\mu\nu} &=& (\rho_{\rm f} c^2 + \epsilon_{\rm f}) u^\mu u^\nu + p_{\rm f} \hat{\eta}^{\mu\nu}
    - 2 \, T_{\rm f} \, \frac{\partial s_{\rm f}}{\partial \alpha_{\mu\lambda}} \,
    ( {\alpha_\lambda}^\nu - {\eta_\lambda}^\nu )
    \nonumber \\
    && - \, T_{\rm f} \frac{\partial s_{\rm f}}{\partial \hat{\omega}_\mu} \hat{\omega}^\nu
    + T_{\rm f}^2 \left( \frac{\partial s_{\rm f}}{\partial \hat{\omega}_\mu} u^\nu
    + u^\mu \frac{\partial s_{\rm f}}{\partial \hat{\omega}_\nu} \right) ,
\label{Tmunudef}
\end{eqnarray}
where $\pRF$ is the pressure in the local rest frame. From a physical point of view, $u_\mu T^{\mu\nu} u_\nu = \rhoRF c^2 + \epsRF$ can be interpreted as the total internal energy density, with the first term on the right-hand side associated with the mass density and the second term associated with the thermodynamic state of the system. The dependence of the energy-momentum tensor on the form of the entropy function is a characteristic feature of nonequilibrium thermodynamics.

The associated entropy equation has the balance form, with the time derivative and flux (\ie flux four-vector) of entropy at the left-hand side, and the entropy production at the right-hand side and it reads
\beq
\begin{split}
\partial_\mu \biggl(\sRF u^\mu + \TRF \pder{\sRF}{\hat{\omega}_\mu} \biggr) =
& - \pder{\sRF}{\alpha_{\mu\nu}} \biggl(\frac{1}{c\lambda_0} \bar{\alpha}_{\mu\nu} + \frac{1}{c \lambda_2}\mathring{\alpha}_{\mu\nu} \biggr) \\ & - \frac{1}{c\lambda_1} \pder{\sRF}{\hat{\omega}_\mu} \hat{\omega}_\mu.
\end{split}
\label{Eq: Entropy balance eq}
\eeq
The entropy-flux four-vector has a contribution proportional to $\hat{\omega}_\mu$, showing that $\hat{\omega}_\mu$ is related to a heat flux. Thus, \Eq{Omega constraint} and \Eq{Entropy balance eq} help to clarify the physical meaning of the four-vector $\omega_\nu$, whose temporal component in the local rest frame represents the temperature, while the spatial components represent the heat flux.

For our stability analysis, we need a particular functional form of the entropy function. We choose the quadratic function
\beq
\begin{split}
\sRF (\rhoRF,\epsRF,\alpha_{\mu\nu},\hat{\omega}_\mu) = & \sRF^\textrm{eq}(\rhoRF,\epsRF) \\
& - \frac{1}{2}\rhoRF [H_\alpha(\alpha_{\mu\nu} \alpha^{\mu\nu} - 1) + H_\omega \hat{\omega}_\mu \hat{\omega}^\mu],
\end{split}
\label{Eq: Quadratic entropy density}
\eeq
where $H_\alpha$ and $H_\omega$ are positive constants, to guarantee the concavity of the entropy with respect to the variables it depends on. A quadratic entropy function is not only consistent with the second-order Israel-Stewart model and the standard formulation of EIT, but should also be sufficient for producing the linearized equations for our stability analysis. For the entropy function (\ref{Eq: Quadratic entropy density}), the energy-momentum tensor (\ref{Tmunudef}) becomes
\beq
\begin{split}
T^{\mu\nu} = & (\rhoRF c^2 + \epsRF)u^\mu u^\nu + \pRF\hat{\eta}^{\mu\nu} \\
& + \rhoRF \TRF\biggl[2H_\alpha(\alpha^{\mu\lambda} \alpha_\lambda\,^\nu  - \alpha^{\mu\nu}) + H_\omega\hat{\omega}^\mu \hat{\omega}^\nu \biggr] \\
& - \rhoRF \TRF H_\omega (\hat{\omega}^\mu u^\nu + u^\mu\hat{\omega}^\nu).
\end{split}
\label{Eq: Energy-momentum tensor}
\eeq
Note that this tensor $T^{\mu\nu}$ is manifestly symmetric.

For the linear stability analysis, we choose to take $\rhoRF$ and $\epsRF$ as independent thermodynamic variables. All thermodynamic properties are expressed in terms of three quantities: adiabatic thermal expansivity $\alpha_s$, isothermal speed of sound $c_T$ and heat capacity ratio $\gamma$. The following expressions hold \cite{hco05_BOOK}

\beq
\rhoRF \pder{\TRF(\rhoRF,\epsRF)}{\rhoRF} + (\epsRF + \pRF) \pder{\TRF(\rhoRF, \epsRF)}{\epsRF} = \frac{1}{\alpha_s},
\label{Eq: Thermodynamic relations, 1}
\eeq

\beq
\pder{\TRF(\rhoRF,\epsRF)}{\epsRF} = \frac{1}{(\gamma - 1) \rhoRF \TRF \alpha_s^2 c_T},
\label{Eq: Thermodynamic relations, 2}
\eeq

\beq
d\pRF = c_T^2 d\rhoRF + (\gamma - 1)\rhoRF c_T^2 \alpha_s d\TRF.
\label{Eq: Thermodynamic relations, 3}
\eeq

Summarizing, our model consists of 14 equations: the continuity equation \Eq{Continuity eq}, the energy-momentum tensor conservation equation \Eq{Energy-momentum conservation eq}, the relaxation equations of the structural variables $\alpha_{\mu\nu}$ and $\omega_\mu$, respectively \Eq{Relaxation alpha} and \Eq{Relaxation omega}. The energy-momentum tensor is expressed as Eq.~(\ref{Tmunudef}). A thermodynamically admissible entropy balance equation \Eq{Entropy balance eq} automatically emerges. To close the model, two state equations are required, relating the independent thermodynamic variables $\epsRF$, $\rhoRF$ to $\TRF$ and $\pRF$ and providing the relations in \Eqs{Thermodynamic relations, 1}, (\ref{Eq: Thermodynamic relations, 2}) and (\ref{Eq: Thermodynamic relations, 3}).
The unknown variables are 14 in total: $\rhoRF$, $\epsRF$, $u_\mu$, $\alpha_{\mu\nu}$ and $\omega_\mu$ (1 + 1 + 3 + 6 + 3 = 14), given the symmetry of the $\alpha_{\mu\nu}$ and the constraints in Eqs.~(\ref{Eq: Velocity constraint}), (\ref{Eq: Alpha constraint}), (\ref{Eq: Omega constraint}).

\section{Linear stability analysis}
\label{sec:Stability analysis}

We here formulate a linear stability analysis for the above-mentioned model of relativistic fluid dynamics. Following standard procedures \cite{HiscockLindblom83,HiscockLindblom85,van08,pu10}, we express a generic perturbation of any variable $q$ from the equilibrium rest state $q_0$ in the form $\delta q = \delta\tilde{q} e^{\omega t + i k x} $, with $t$ the time and $x$ one of the space directions in the rest frame.

\subsection{Equilibrium rest-frame}

We consider an equilibrium state corresponding to the rest state. The values of the variables in such a state are

\beq
\begin{split}
\rho^\equi               & = \rho_{\textrm{f},0}  ,\,\,\,\,  p^\equi = p_{\textrm{f},0}   ,\,\,\,\,  T^\equi = T_{\textrm{f},0},  \\
(u^0)^\equi              & = 1                    ,\,\,\,\,  (u^{i})^\equi  = 0           ,\,\,\,\,  (\omega^0)^\equi  = -T_{\textrm{f},0}   ,\,\,\,\,  (\omega^i)^\equi  = 0,  \\
(\alpha^{00})^\equi      & = - 1                  ,\,\,\,\,  (\alpha^{0i})^\equi = (\alpha^{i0})^\equi = (\alpha^{ij})^\equi = 0,      \\
(\hat{\eta}^{00})^\equi  & = (\hat{\eta}^{0i})^\equi = (\hat{\eta}^{i0})^\equi = 0        ,\,\,\,\,  (\hat{\eta}^{ij})^\equi  = \delta_{ij}
\end{split}
\label{Eq: Rest frame}
\eeq
with constant values $\rho_{\textrm{f},0}$, $T_{\textrm{f},0}$ and $p_{\textrm{f},0}$ for a particular equilibrium state.\\
The first order variations of the variables, namely the unknowns of the linearized problem, are
$\delta\rho, \delta T, \delta u^x, \delta u^y, \delta u^z, \delta\omega^x, \delta\omega^y, \delta\omega^z, \delta\alpha^{xx}, \delta \alpha^{xy}, \delta \alpha^{xz}, \delta \alpha^{yy}$, $\delta \alpha^{yz}, \delta \alpha^{zz}$. The following relationships hold

\beq
\begin{split}
\delta u^0             & = 0    ,\,\,\,\,  \delta \omega^0       =  -\delta T                             ,\,\,\,\,   \delta\omega^i = \delta\omega_i, \\
\delta\alpha^{00}      & = 0    ,\,\,\,\,  \delta\alpha^{0i}     = \delta\alpha^{i0}     = -\delta u^i    ,\,\,\,\,   \delta\alpha^{ij} = \delta\alpha^{ji}, \\
\delta\hat{\eta}^{00}  & = 0    ,\,\,\,\,  \delta\hat{\eta}^{i0} = \delta\hat{\eta}^{0i} = -\delta u^i.   \\
\end{split}
\label{Eq: 1st order variations}
\eeq

\subsection{Linearized equations}

To prepare the stability analysis, we write the linearized evolution equations of the model as

\begin{align}%
& \partial_0\rhoRF  = -\rhoRF \partial_i u_i,
\label{Eq: Simplified model, eq 1} \\
& \partial_0 \epsRF = -(\epsRF + \pRF - \rhoRF\TRF^3H_\omega)\partial_i u_i + \rhoRF\TRF^2H_\omega\partial_i\omega_i,
\label{Eq: Simplified model, eq 2} \\
& (\rhoRF c^2 + \epsRF + \pRF - \rhoRF \TRF^3 H_\omega) \partial_0 u_j = \label{Eq: Simplified model, eq 3} \\
& \qquad \qquad -\partial_j \pRF + 2 \rhoRF\TRF H_\alpha \partial_k \alpha_{jk} + \rhoRF \TRF^2H_\omega \partial_0 \omega_j,  \nonumber \\
& \partial_0 \omega_j = \partial_j \omega_0 - \frac{1}{c\lambda_1}(\omega_j + u_j\omega_0),
\label{Eq: Simplified model, eq 4} \\
& \partial_0 \alpha_{jk} = \partial_j u_k + \partial_k u_j  - \frac{1}{c\lambda_2} \alpha_{jk} - \biggl(\frac{1}{c\lambda_0} - \frac{1}{c\lambda_2} \biggr) \frac{\alpha_{ll}}{3} \delta_{jk},
\label{Eq: Simplified model, eq 5}
\end{align}
where the subscript $0$ has been dropped in $T_{\textrm{f},0}$, $\rho_{\textrm{f},0}$ and $\epsilon_{\textrm{f},0}$ for simplicity of notation. We define the following parameters

\beq
X = \frac{2\TRF H_\alpha}{c_T^2}, \quad Y = \frac{\TRF^2 H_\omega}{(\gamma - 1) \alpha_{s}c_T^2}, \quad Z = (\gamma - 1)\TRF \alpha_s.
\label{Eq: Parameters}
\eeq

By making use of the thermodynamic relations Eqs.~(\ref{Eq: Thermodynamic relations, 1}),
(\ref{Eq: Thermodynamic relations, 2}), (\ref{Eq: Thermodynamic relations, 3}), from \Eq{Simplified model, eq 2} we derive the temperature equation

\beq
\partial_0\TRF = - \frac{1 - Y}{\alpha_s} \partial_i u_i + \frac{Y}{\TRF\alpha_s}\partial_i \omega_i,
\label{Eq: Temperature eq}
\eeq
and, from \Eq{Simplified model, eq 3}, the equation of motion

\beq
\partial_0 u_j = Q \biggl(-\frac{1}{\rhoRF}\partial_j\rhoRF - \frac{Z}{\TRF}\partial_j\TRF + X \partial_k\alpha_{jk} + \frac{YZ}{\TRF}\partial_0 \omega_j \biggr),
\label{Eq: Eq of motion}
\eeq
where
\beq
Q = \frac{\rhoRF c_T^2}{\rhoRF c^2 + \epsRF + \pRF - YZ\rhoRF c_T^2}.
\label{Eq: Parameter Q}
\eeq

Natural ranges of parameters are: $1 < \gamma < 2$ , $X > 0$, $0 < Y < 1$, $Q > 0$, $ \lambda_j > 0$.
The constraints on $Y$ arise from \Eq{Temperature eq}: the temperature should increase when
the system is compressed ($\partial_i u_i < 0$), hence $Y < 1$.
From the physical point of view, $Q$ gives an indication on the relativistic character of a fluid.
The limit $Q \rightarrow 0$ corresponds to a nonrelativistic fluid. As the speed of sound $c_T$
approaches the speed of light $c$, the parameter $Q$ grows.
The time scales $\lambda_0$ and $\lambda_2$ can be
chosen independently, without the need to impose any further inequality, as it can be seen by splitting
the tensorial relaxation equation (\ref{Eq: Simplified model, eq 5}) for the $\alpha_{jk}$ into its tracefree and trace part

\begin{align}
& \partial_0 \mathring{\alpha}_{jk} = \partial_ju_k + \partial_k u_j -\frac{2}{3}\partial_i u_i \delta_{jk} - \frac{1}{c\lambda_2} \mathring{\alpha}_{jk},
\label{Eq: Splitting relaxation alpha, tracefree} \\
& \partial_0 \alpha_{ii} = 2\partial_i u_i - \frac{1}{c\lambda_0}\alpha_{ii}.
\label{Eq: Splitting relaxation alpha, trace}
\end{align}

For small relaxation times $\lambda_j$, the quantities $\mathring{\alpha}_{jk}$, $\alpha_{ii}$ and $\omega_j$ relax quickly to the quasi-stationary solutions; hence, from \Eqs{Splitting relaxation alpha, tracefree}, \eq{Splitting relaxation alpha, trace} and \eq{Simplified model, eq 4} respectively, one obtains

\begin{align}
\mathring{\alpha}_{jk} & = c\lambda_2\biggl(\partial_j u_k + \partial_k u_j -\frac{2}{3}\partial_i u_i \delta_{jk} \biggr),
\label{Eq: Quasi-stationary solution, alpha tracefree} \\
\alpha_{ii} & = 2c\lambda_0\partial_i u_i,
\label{Eq: Quasi-stationary solution, alpha trace} \\ \nonumber \\
-\omega_j & = u_j \TRF - c\lambda_1\partial_j\TRF.
\label{Eq: Quasi-stationary solution, omega}
\end{align}
These equations lead to the proper interpretation of the relaxing structural variables $\alpha_{jk}$ and $\omega_j$: Eqs.~(\ref{Eq: Quasi-stationary solution, alpha tracefree}) and
 (\ref{Eq: Quasi-stationary solution, alpha trace}) contain the proper combinations of the velocity gradient tensor for the Newtonian shear and dilatational stresses expected in the nonrelativistic limit; \Eq{Quasi-stationary solution, omega} contains the heat flux (multiplied by a factor). In particular, $-\omega_j$ consists of the convective and conductive flux contributions in the temperature equation.

\subsection{Block structure}

%
%

The linearized system can be rewritten in the matrix form $\textbf{M} \cdot \delta q = 0$, where $\textbf{M}$ is a $14\times14$ matrix with block-structure (6 + 3 + 3 + 1 + 1) and
$\delta q = (\delta \rho, \delta T, \delta u^x, \delta \omega^x, \delta \alpha^{xx}, \frac{\delta \alpha^{yy} + \delta \alpha^{zz}}{2}, \delta u^y,\delta \omega^y,\delta \alpha^{xy}, \delta u^z,\\ \delta\omega^z,\delta\alpha^{xz},\delta \alpha^{yy}-\delta \alpha^{zz},\delta \alpha^{yz})^T$. Without loss of generality, we can choose $\rhoRF = \TRF = c = \lambda_1 = 1$, which is equivalent to choosing the units for density, temperature, velocity and time. The matrix $\textbf{M}$ is

\beq
\textbf{M} =  \begin{bmatrix}
  \textbf{N} & 0 & 0 & 0 & 0 \\
  0 & \textbf{R} & 0 & 0 & 0\\
  0 & 0 & \textbf{R} & 0 & 0\\
  0 & 0 & 0 & f & 0 \\
  0 & 0 & 0 & 0 & f
\end{bmatrix},
\label{Eq: Linearized system, block form}
\eeq
with $f = [1\times1]$, $\textbf{R} = [3\times3]$ and $\textbf{N}=[6\times6]$. In particular

\beq
f = \Phi + 3F,
\label{Eq: Block f, 1x1}
\eeq

\beq
\textbf{R} =  \begin{bmatrix}
  \Phi   & -YZ\Phi      & -XQ \mi     \\
  3G     & \Phi + 3G    & 0           \\
  -\mi   & 0            & \Phi + 3F   \\
\end{bmatrix},\label{Eq: Block R, 3x3}
\eeq

\beq
\small
\setlength{\arraycolsep}{0.2pt}
\renewcommand{\arraystretch}{1.2}
\textbf{N} =  \begin{bmatrix}
  \Phi  & 0      & \mi               & 0                  & 0          & 0         \\
  0     & \Phi   & \frac{1-Y}{A}\mi  & -\frac{Y}{A}\mi    & 0          & 0         \\
  Q\mi  & \,\,QZ\mi  & \Phi          & -YZQ\Phi           & -XQ\mi     & 0         \\
  0     & -\mi   & 3G                & \Phi + 3G          & 0          & 0         \\
  0     & 0      & -2\mi             & 0                  & \Phi+E+2F  & 2(E - F)  \\
  0     & 0      & 0                 & 0                  & E - F      & \Phi+2E+F \\
\end{bmatrix}.
\label{Eq: Block N, 6x6}
\eeq
In the above equations, we defined $\Phi = \omega / k$ as well as the additional dimensionless parameters

\beq
A = \alpha_s \TRF, \,\,\,\,\, E = \frac{1}{3k\lambda_0}, \,\,\,\,\, F = \frac{1}{3k\lambda_2}, \,\,\,\,\, G = \frac{1}{3k\lambda_1}.
\label{Eq: Parameters A, E, F, G}
\eeq

To assess the stability, we look for the roots $\Phi_n$ of the characteristic polynomial $\det\textbf{M} = 0$, with $n = 1,...,14$, \ie we look for the values of $\Phi$ for which the system has nontrivial solutions for any initial perturbation of the variables $\delta \tilde{q}_n$ respect to the equilibrium rest state $q_0$. We assume $k \epsilon \mathbb{R}^+$, therefore, if $\Re(\Phi_n) < 0$ also $\omega_n<0$, with $\omega_n = k\Phi_n$; hence any initial perturbations $\delta \tilde{q}_n$ will decay in time and the system will be stable.

Given the block structure of the matrix \textbf{M} the characteristic polynomial can be factorized as
\beq
\det\textbf{M} = \det\textbf{N} \cdot (\det\textbf{R})^2 \cdot f^2 = 0.
\label{Eq: Determinant block matrix M}
\eeq
We consider the different blocks separately. The solutions of the characteristic polynomial \Eq{Determinant block matrix M} are found for
\begin{align}
f = & \, \Phi + 3F = 0,
\label{Eq: Solutions characteristic polynomial, 1eq} \\
\det\textbf{R} = & \, 3 G Q X + \Phi (9 F G + Q X + 9 F G Y Z) \label{Eq: Solutions characteristic polynomial, 2eq} \\
& + \Phi^2 (3 F + 3 G + 3 G Y Z) + \Phi^3 = 0, \nonumber \\
\det\textbf{N} = & \, 0.
\label{Eq: Solutions characteristic polynomial, 3eq}
\end{align}
\Eq{Solutions characteristic polynomial, 1eq} implies $\Phi = -3F$ , which is always real and negative, thus corresponding to decaying perturbations. In particular, it corresponds to $\omega =  - 1 / \lambda_2$, \ie a constant decaying rate related to the characteristic relaxation constant of $\mathring{\alpha}_{\mu \nu}$ and independent of $k$, the spatial wavelength of the perturbation. This condition characterizes the relaxation of shear and second normal stress difference.

In the physically acceptable range of parameters, \Eq{Solutions characteristic polynomial, 2eq} has only solutions featuring a negative real part. To proof this, we apply the Routh-Hurwiz criterion \cite{hur95} for a $3^{\textrm{rd}}$ order polynomial $a_0 + a_1 \Phi + a_2 \Phi^2 + \Phi^3 $. According to this criterion, such a polynomial has only solutions with a negative real part if $a_i > 0$ for $\forall i$ and $a_1 a_2 > a_0$. The first condition can be verified from \Eq{Solutions characteristic polynomial, 2eq}, given the physically acceptable ranges of the parameters, which must all be real and positive. The second condition can be rewritten as

\begin{align}
a_2 a_1 - a_0 & = 27 F G (F + G + FYZ + 3GYZ + G Y^2 Z^2)  \label{Eq: Routh-Hurwitz, det R = 0} \\
              & \,\,\, + 3 Q X (F + G Y Z) > 0 ,  \nonumber
\end{align}
which is true for real valued, positive $F$, $G$, $X$, $Y$, $Z$.
In the limits of $Z = 0$ or $Y = 0$, corresponding to the limit of $T_f = 0$ (since $\gamma = 1$ and $\alpha_s = 0$ would provide unphysical values of $Y$, for nonzero values of $H_\omega$), one solution is the purely real negative $\Phi = -3F$, corresponding to an additional shear mode $\omega =  - 1 / \lambda_2$.
The same would happen in the limit case $Q = 0$ (nonrelativistic limit) or in the case that $F = G$, \ie $\lambda_1 = \lambda_2$. In particular in the nonrelativistic limit case of $Q = 0$, additional roots of \Eq{Solutions characteristic polynomial, 2eq} will be $\Phi = -3G (1 + YZ)$ and $\Phi = 0$. The latter corresponds to a case of marginal stability, while the former corresponds to a perturbation decaying in time with $\omega = - (1 + T_f^3H_\omega/c_T)/\lambda_1$, independent of $k$. This can be interpreted as a non-hydrodynamic mode, where the decaying occurs faster at higher temperatures. Physically, one can imagine a higher thermal agitation of the fluid particles to correspond to a higher dissipation rate.

We now consider \Eq{Solutions characteristic polynomial, 3eq}. To determine the sign of the roots, we rearrange \Eq{Solutions characteristic polynomial, 3eq} as

\begin{widetext}
\begin{equation}
X = -\frac{(3E + \Phi) (3F + \Phi) \{(\Phi^2 + Q) [A \Phi (3G + \Phi) + Y] +
      \Phi Q [\Phi (-1 + Y)^2 + 3G (1 + A \Phi^2 Y)] Z\}}{2 \Phi (2E + F +
      \Phi) Q [A \Phi (3G + \Phi) + Y]},
\label{Eq: Det N = 0, cdt 1}
\end{equation}
\end{widetext}
which can be compactly rewritten as $ X = - N/D$ with $N$ the numerator and $D$ the denominator of the fraction.
In the case of $\Phi$ purely real, positive values of $\Phi$ would lead to both $N>0$ and $D>0$, hence $X < 0$. Since these are physically unacceptable values of $X$, purely real positive roots are not possible and purely real roots $\Phi$ can only be negative.
Purely imaginary values of $\Phi$ correspond to an oscillatory non-growing solution of the linear system and do not concern its stability.
Therefore further investigation is required only for the case of a complex-valued root,
\beq
\Phi = R + \mi I,
\label{Eq: Phi complex}
\eeq
with $R = \Re(\Phi) \neq 0$ and $I = \Im(\Phi) \neq 0$. We substitute \Eq{Phi complex} inside \Eq{Det N = 0, cdt 1} and we impose $\Im(X) = 0$, since we know from \Eq{Parameters} that $X$ has to be a real number. The condition for such a requirement to be met is expressed in terms of $Q = Q(R,I,E,F,G,Y,Z)$ and inserted into \Eq{Det N = 0, cdt 1}, thus finding

\begin{widetext}
\beq
X = - \frac{ [I^2 + (3E + R)^2][I^2 + (3F + R)^2] T_1}{2|\Phi|^2 \{6E^2 (3F + 2R) + 2(FI^2 + 3F^2R + R|\Phi|^2 + 3FR^2 ) + EI^2 + 9(F + R)^2]\}T_2},
\label{Eq: Re(X), 1}
\eeq
\end{widetext}
with
\begin{align}
T_1 & = 2A R \{ A [ 2I^2 R^2 + R^4 + 3G(2R + 3G) |\Phi|^2 ]       \label{Eq: T1}  \\
    &\,\,\,\,  +  2R(3G + R)Y \} + 2R(A I^2 - Y)^2                     \nonumber  \\
    &\,\,\,\,  +  3GZ|\Phi|^2 [2AR (3G + 2R) + Y (A|\Phi|^2 - 1)^2]    \nonumber  \\
    &\,\,\,\,  +  Z|\Phi|^2 [ 2AR |\Phi|^2 (1-Y)^2 ],                  \nonumber  \\
    \nonumber \\
\end{align}
\begin{align}
T_2 & = (A I^2 - Y)^2 + A \{A [2 I^2 R^2 + R^4 + 3 G |\Phi|^2 (3 G + 2 R)] \nonumber  \\
    &\,\,\,\,  + 2 R (3 G + R) Y \}.
\label{Eq: T2}
\end{align}
We compactly rewrite \Eq{Re(X), 1} as $ X = - N'/D'$, where $N'$ and $D'$ are the numerator and the denominator of the fraction, respectively.
From the equations above, one can see that $T_1 > 0$ and $T_2 > 0$, in the case of $R > 0$. Thus $N' > 0$, $D' > 0$ and $X < 0$, which is against the assumptions on the acceptable values of $X$. Therefore, it has to be $R = \Re(\Phi) \leq 0$ and the proof of the unconditional stability of the system is complete, within the physically acceptable range of values of the parameters.

In \Figs{omega(k), Q = 0.001}-\ref{Fig:omega(k), Q = 1} we show the real and imaginary part of the $\omega$ as a function of $k$, for different values of the parameter $Q$: $Q =0.001$, $Q = 0.1$ and $Q = 1$. The other parameters are: $X = 1$, $Y = 1/2$, $Z = 1$, $A = 1/3$, $\lambda_0 = 1$, $\lambda_1 = 2/3$ and $\lambda_2 = 1/6$.
The solid lines represent the eigenvalues derived from the $6\times6$ block $\textbf{N}$ (see \Eq{Solutions characteristic polynomial, 3eq}). The dashed lines represent the eigenvalues derived from the $3\times3$ block $\textbf{R}$ (see \Eq{Solutions characteristic polynomial, 2eq}). Since there are two identical such blocks in the matrix $\textbf{M}$, the multiplicity of each dashed line has to be considered double. The colours (online) in the figures are redundant with the numbering, to facilitate the readability.

In the case of $Q = 0.001$, close to the nonrelativistic limit, we find several modes that have been previously identified by other authors \cite{pu10}: the line labeled $1$ (green online) is characterized by a proportionality to $-k^2$ near $k=0$ and identifies a diffusive mode for $k < k_c$, where $k_c$ is a characteristic wavelength. The line $2$ (brown online), represents a relaxation mode related to the heat fluxes $\omega_j$: this can be seen from the fact that, for $k = 0$, it takes the value $\omega = -1/\lambda_1$ (see \Eq{Simplified model, eq 4}). The real parts of $1$ and $2$ merge for $k > k_c$, while the complex parts becomes nonzero and conjugate, turning both $1$ and $2$ into two non-hydrodynamic modes. The modes $3$ and $4$ (orange online) are complex conjugate with negative real part, hence they represent two sound modes. The modes $5$ (magenta online) and $6$ (yellow online) have negative real part, reaching an asymptote as $k$ increases and null complex part, thus they can be identified as two non-propagating relaxation modes. In particular, at $k = 0$, the mode $5$ takes the value $\omega = -1/\lambda_0$ while the mode $6$ takes the value $\omega = -1/\lambda_2$. Hence we conclude that the mode $5$ is related to the relaxation of $\alpha_{ii}$ (see \Eq{Splitting relaxation alpha, trace}) and the mode $6$ is related to the relaxation of $\mathring{\alpha}_{jk}$ (see \Eq{Splitting relaxation alpha, tracefree}).
From the $3\times3$ block $\textbf{R}$, three modes appear, represented by the dashed lines: two shear modes, which turn into non-hydrodynamic modes for $k$ larger than a characteristic value $k_c'$ (dark-grey and light-grey, blue and red online) and one non-hydrodynamic mode (black). For $Q = 1$, the lines $1$ and $2$ represent two shear modes for $k < k_c$, since they are shifted downwards respect to the origin of the axes, while the other modes retain a similar meaning. In the case of $Q = 0.1$, an intermediate situation between the two previous cases appears and the different modes are combined in a nontrivial way.

\begin{figure*}
\centering
\includegraphics[width=1\textwidth]{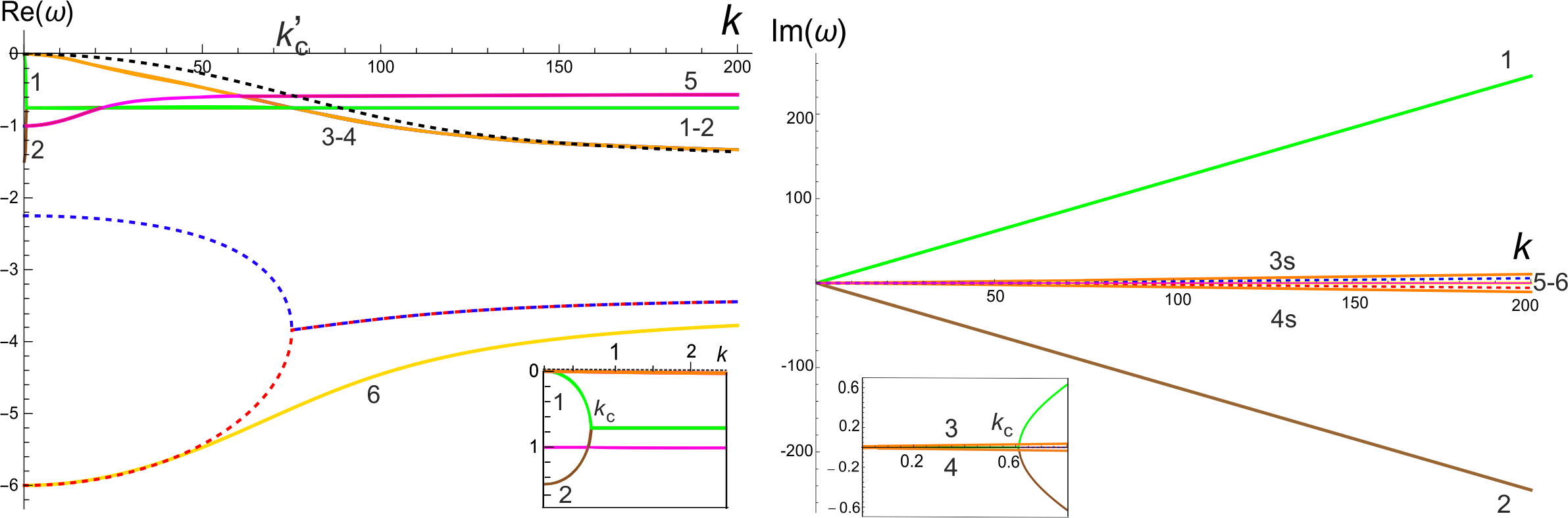}
\caption{Real (left) and imaginary part (right) of the eigenmodes $\omega$ as a function of $k$, for $Q = 0.001$. Other parameters: $A = 1/3$, $X = 1$, $Y = 1/2$, $Z = 1$, $\lambda_0 = 1$, $\lambda_1 = 2/3$ and $\lambda_2 = 1/6$. The dashed lines represent the roots of the characteristic polynomial of the $3\times3$ block $\textbf{R}$; the solid lines represent the roots of the characteristic polynomial of the $6\times6$ block $\textbf{N}$. Both $\omega$ and $k$ are depicted in nondimensional units.}
 \label{Fig:omega(k), Q = 0.001}
\end{figure*}

\begin{figure*}
\centering
\includegraphics[width=1\textwidth]{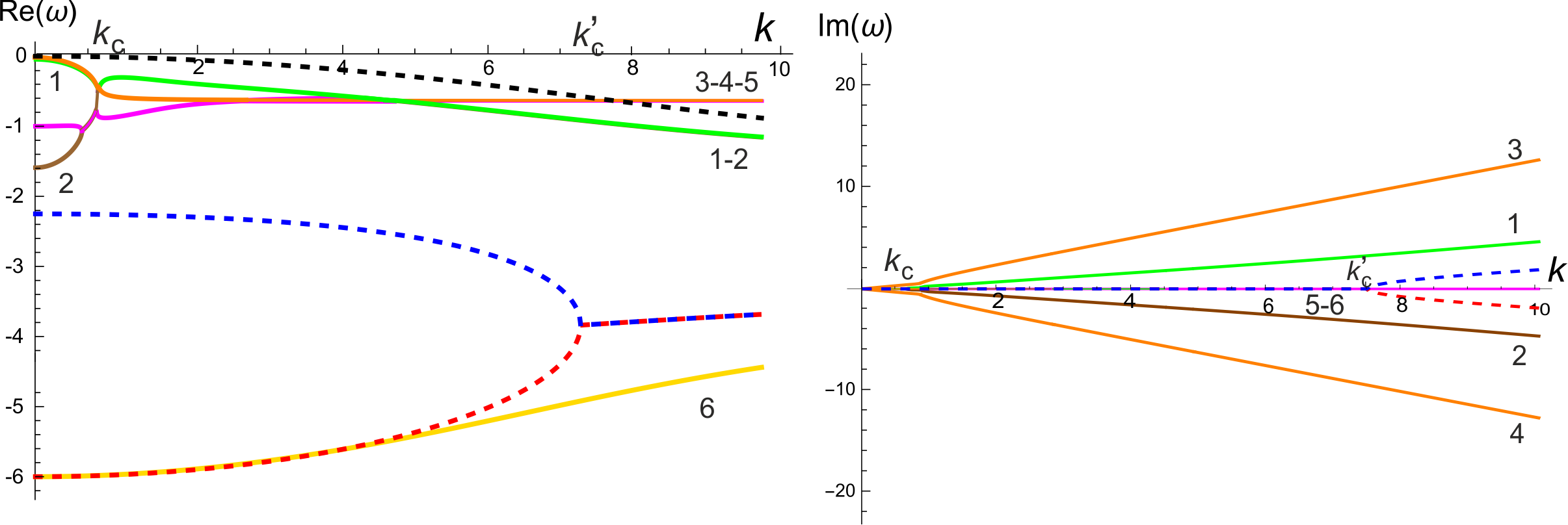}
\caption{Real (left) and imaginary part of the eigenmodes $\omega$ as a function of $k$, for $Q = 0.1$. Other parameters and line style code: see \Fig{omega(k), Q = 0.001}.}
 \label{Fig:omega(k), Q = 0.1}
\end{figure*}

\begin{figure*}
\centering
\includegraphics[width=1\textwidth]{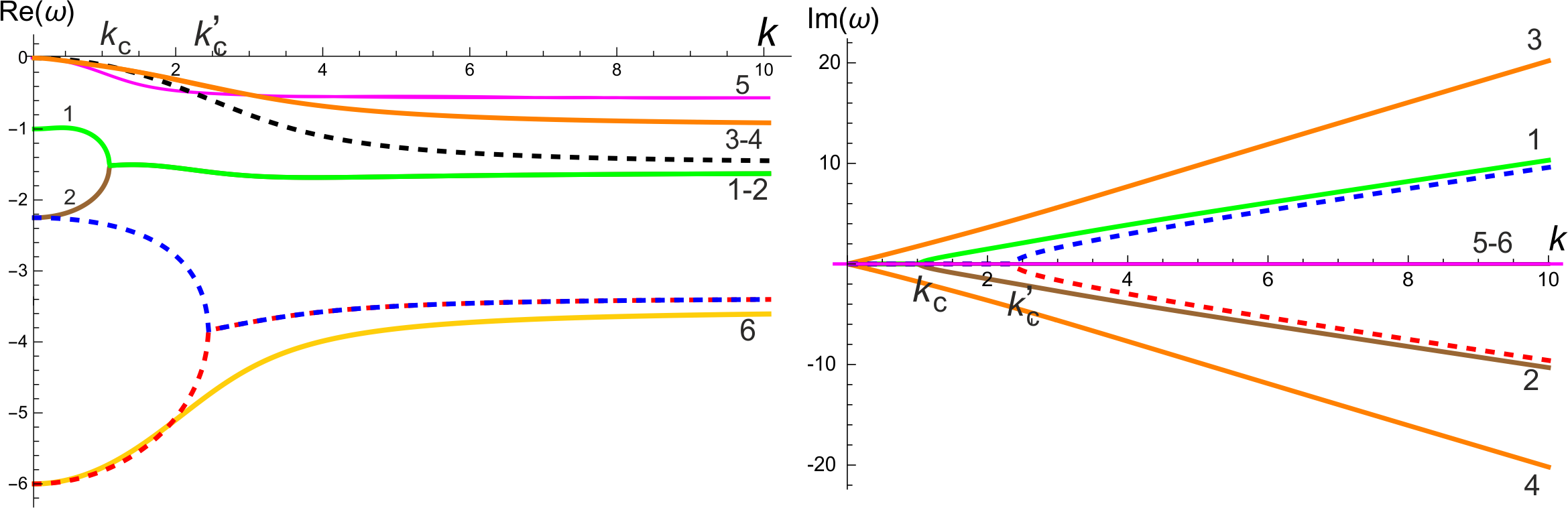}
\caption{Real (left) and imaginary part of the eigenmodes $\omega$ as a function of $k$, for $Q = 1$. Other parameters and line style code: see \Fig{omega(k), Q = 0.001}.}
 \label{Fig:omega(k), Q = 1}
\end{figure*}

\section{Summary and discussion}
\label{sec:Summary}
In the present paper we considered the thermodynamically consistent model for relativistic fluid dynamics developed within the GENERIC framework \cite{hco98,hco98b}, where the treatment of nonequilibrium phenomena (momentum and heat transport) is described by means of additional structural variables to prevent the problem of the infinite propagation speeds. We performed a linear stability analysis of the equilibrium rest state and we analytically proved that, in the entire range of the physically meaningful parameters, the model is unconditionally stable. We verified that, in the nonrelativistic limit, the standard shear, sound, diffusive and non-propagating modes are retrieved while, as relativistic effects become more relevant, the modes start to mix with each other in a nontrivial way.

In nonrelativistic fluid dynamics, the equations for liquids and gases possess the same Navier-Stokes-Fourier form and differ only by the characteristic values of the transport coefficients. This is a consequence of the fact that the basic equations express the balance laws for the conserved quantities mass, momentum and energy in a straightforward form. In relativistic fluid dynamics, we are faced with additional relaxation equations for non-conserved structural variables so that there is no reason to expect universal equations that encompass both liquids and gases.

The GENERIC equations of relativistic fluid dynamics include a Poisson-bracket formulation of the convection mechanism. A similar Poisson bracket has not been found for the 13- or 14-moment equations derived from the nonrelativistic or relativistic Boltzmann equation. The problem is that the convection terms for increasingly higher moments in momentum space are strongly coupled and hence very complicated so that they become difficult, if not impossible, to truncate. The situation is very different if the structural variables describe features related to forces and differences in position space. This is the reason why we expect a serious difference between fluid dynamics of relativistic liquids and gases. The heuristic derivation of the GENERIC equations based on the knowledge of complex fluids in \cite{hco18_BOOK} underlines their liquid-like background, whereas the Israel-Stewart model and the EIT equations are based on moment expansions for the Boltzmann equation and hence are appropriate for gas-like fluids. Note that, in the case of quark-gluon plasmas, recent experimental findings showed that a liquid-like description may be more accurate \cite{STA17}. For gas-like fluids, GENERIC structure has so far been found only in the Boltzmann equation or the equivalent infinite moment hierarchy.

With the present paper we performed one step forward in showing that the thermodynamically consistent model developed within the GENERIC framework \cite{hco98,hco98b} should be considered as a valuable option, a worthwhile alternative to the existing ones, when addressing the study of relativistic fluids. To gain full confidence, linear stability should also be investigated in a boosted system \cite{pu10}, which is considerably more complicated and hence beyond the scope of the present analytical study (more terms in evolution equations, angle between boost and wave vector of perturbation). The elegance associated with the Poissonian structure of the convection terms in the relaxation equations (\ref{Eq: Relaxation alpha}) and (\ref{Eq: Relaxation omega}) for the structural variables is an important argument in favor of the equations derived within the GENERIC framework. More physical entropy functions, presumably containing logarithmic rather than quadratic contributions, open the door to more realistic modeling, where the form (\ref{Tmunudef}) of the energy-momentum tensor is prescribed by thermodynamic admissibility.

\begin{acknowledgments}
The authors would like to express their particular gratitude to Martin Kr\"{o}ger, who greatly contributed to the final proof of the stability of the problem.
They would also like to thank Alberto Montefusco, Martin Callies and Michele Graf for insightful discussions.
\end{acknowledgments}

%

\end{document}